\documentclass[twocolumn]{aastex631}
\usepackage{longtable}
\usepackage{subfigure}
\usepackage{url}
\usepackage{threeparttable}
\usepackage{chemformula}
\usepackage{txfonts}
\usepackage{CJK}
\usepackage{bm}

\newcommand{\egcite}[2]{(e.g., \citet{#1}, see Section~#2)}
\begin{document}
\begin{CJK}{UTF8}{gbsn}

\title{Expansion and Spectral Softening of the Dust Scattering Rings of GRB 221009A}

\author{Guoying Zhao（赵国英）}
\thanks{zhaogy28@mail2.sysu.edu.cn} 
\affiliation{School of Physics and Astronomy, Sun Yat-Sen University, Zhuhai, 519082, P. R. China}
\affiliation{CSST Science Center for the Guangdong-Hongkong-Macau Greater Bay Area, Sun Yat-Sen University, Zhuhai, 519082, P. R. China}
\author{Rong-Feng Shen（申荣锋）}
\thanks{shenrf3@mail.sysu.edu.cn}
\affiliation{School of Physics and Astronomy, Sun Yat-Sen University, Zhuhai, 519082, P. R. China}
\affiliation{CSST Science Center for the Guangdong-Hongkong-Macau Greater Bay Area, Sun Yat-Sen University, Zhuhai, 519082, P. R. China}


\begin{abstract}
Expanding X-ray halo or rings appear when short pulses of X-ray radiation from a background source are scattered by clouds of dust in the Milky Way. We study the X-ray rings of the brightest gamma-ray burst (GRB) 221009A, detected by the {\it Swift} X-Ray Telescope. The rings center on the GRB position and their angular radii increase with time. We identify five major expanding rings, and our modeling of their expansion history suggests that they are scattered off, respectively, from five dusty clouds at distances of 0.4-13 kpc from the observer. Given an assumed prompt X-ray fluence of this GRB, the fluxes of those rings suggest that these clouds have dust grain column densities of $10^{7\sim8}~\mathrm{cm^{-2}}$. More interestingly, our time-dependent spectral analysis of these rings show that they all experience spectral softening, i.e., getting softer as they expand, with spectral indices ranging from 2.2 to 5, consistent with what the dust scattering model predicts.
\end{abstract}




\section{Introduction}

X-rays from bright point sources are affected by the scattering of interstellar dust particles as they are directed towards the observer, and the small-angle scattering of X-rays by dust grains can form detectable X-ray `halos' around distant X-ray sources. \cite{1983Natur.302...46R} first confirmed the presence of a faint X-ray halo around the bright X-ray source GX339-4 by analysing the {\it Einstein} Observatory data. \cite{1986ApJ...302..371M} used the {\it Einstein} to observe six such `halo' sources, and they found that the intensities of the halos are strongly correlated with the extinction of visible light or the distance of the X-ray photons through the Galaxy's dust layer. \cite{2000A&A...357L..25P} used the variability time delay between the halo and the source \citep{1973A&A....25..445T} to measure the distance of Cygnus X-3 to be 9 kpc. This measurement was consistent with previous results. \cite{2002ApJ...581..562S} analyzed the halo of X-ray binary GX 13+1 observed by {\it Chandra}and determined the light-of-sight position, size distribution and density of the scattering dust grains. 

Gamma-ray bursts (GRBs) \citep{1973ApJ...182L..85K} are the most violent explosions other than the Big Bang. They are sudden and random, increase in $\gamma-$ray radiation from cosmological distances. GRBs produce high X-ray fluxes for short periods of time, usually $\leq1000$ s. The duration of a GRB in the gamma-ray band is generally characterized by $T_{90}$, which is defined as the duration between $5 \% \sim 95 \%$ of the cumulative flux as a percentage of the total $\gamma-$ray flux. \cite{1993ApJ...413L.101K} shows that GRBs can be classified into long GRBs ( $\gtrsim$ 2 s) and short GRBs ( $\lesssim$ 2 s) based on their duration. Short GRBs are thought to be mergers of binary neutron star systems, whereas long GRBs are produced by collapse of massive stars~\citep{2009ApJ...703.1696Z}.  

The transient nature of the burst, combined with the high initial X-ray flux, means that the dust-scattered X-rays may be visible with relatively high surface brightness. Dust scattering rings have so far only been observed around five GRBs: 031203, 050713A, 050724, 061019, 070129 and 160623A~\citep{2004ApJ...603L...5V,2006A&A...449..203T,2006ApJ...639..323V,2007A&A...473..423V,2017MNRAS.472.1465P}. The study of dust scattering rings helps to derive the distance, composition and grain size distribution of the dusty cloud.

On October 9th, 2022 at 13:16:59.988 UTC, the {\it Fermi} Gamma-ray Burst Monitor (GBM) was triggered by a very bright GRB. It was also detected by the {\it Fermi} Large Area Telescope (LAT) \citep{2023ApJ...952L..42L,2022GCN.32637....1B}, {\it Swift} \citep{2022GCN.32634....1L}, HXMT-{\it Insight}, {\it HEBS} ({\it GECAME}-C)~\citep{2023arXiv230301203A} and {\it LHAASO} reported gamma-ray detections of up to 18 TeV~\citep{2023Sci...380.1390L}. GRB 221009A is the brightest GRB in history, whose prompt emission lasted for approximately 289 s \citep{2023ApJ...952L..42L}. This source is located at RA = 288$^{\circ}$.264, Dec = 19$^{\circ}$.768 \citep{2023ApJ...952L..42L}, close to the Galactic Plane ($p$ = 4$^{\circ}$.32 and $l$ = 52$^{\circ}$.96). The afterglow was observed by the X-shooter on ESO's Very Large Telescope UT3 and a redshift of $z$ = 0.151 was determined \citep{2023arXiv230207891M}, corresponding to a distance of 749.3 Mpc. The close proximity and the huge energy output make this burst an extraordinary event.
 
In this paper, we study the X-ray halo around GRB 221009A observed by {\it Swift}/XRT on six observations shortly after the burst. After we started the work, several other works have appeared.~\cite{2023MNRAS.521.1590V} analysed XRT data from the days following the {\it Swift} detection of GRB 221009A and found a total of 16 rings due to scattering from dust clouds at distances of 3.4 $-$ 14.7 kpc from Earth. \cite{2023ApJ...946L..24W} uesed {\it Swift}/XRT data to find a series of dust clouds that ranged from 0.2 kpc to over 10 kpc from the observer. \cite{2023ApJ...946L..21N} used {\it IXPE} data to estimate the mean distances of the inner and outer rings as 14.41 and 3.75 kpc, respectively, from the temporal evolution of the two dust scattering rings. \cite{2023ApJ...946L..30T} used {\it XMM-Newton} data to find 20 dust scattering rings around the GRB 221009A produced by dust between 0.3 kpc and 18.6 kpc. 

Although the different teams have analyzed the dust scattering rings from the same source, not only are the numbers of dust scattering rings obtained different, but also the distances of the dust clouds from the observer vary. In view of these, our first goal is to determine the location and the grain column densities of the dust clouds by studying the temporal evolution of the dust scattering rings. 

More importantly, it has been long known (since \cite{1965ApJ...141..864O}) that the dust scattering differential cross section has a strong dependence on the photon wavelength, which will lead to significant spectral softening in the scattered flux as predicted in ~\cite{2007ApJ...660.1319S}
 and~\cite{2009MNRAS.393..598S}. However, this feature has not been fully tested in the observational data of dust scattering rings. Therefore, our second goal is to investigate this spectral feature for GRB 221009A. For this reason, our study would be focusing only on the most prominent rings.

The paper is structured as follows. We describe the observed data and the image extraction, in Section \ref{sec:obs}. We analyze the dust scattering rings and model their expansion in Section~\ref{sec:expanding}. The rings' spectral analysis and the modeling of the softening feature are presented in Section ~\ref{sec:SPECTRAL}. Then, we estimate the dust column densities from the ring fluxes in Section \ref{sec:density}. We explain the similarities and differences between our results and other four groups in Section \ref{sec:Comparison}. Finally, we summarize the result and conclude in Section \ref{sec:con}.

\section{X-RAY IMAGE EXTRACTION} \label{sec:obs}
The {\it Swift} Burst Alert Telescope (BAT) was triggered by GRB 221009A on October 9th, 2022 at 14:10:17 UTC~\citep{2022GCN.32632....1D}. The {\it Swift} X-ray telescope (XRT) observations of the GRB 221009A filed started on 14:13:30 UTC, about 3 minutes after the BAT trigger. Actually, the burst triggerd {\it Fermi} and HXMT-{\it insight} even earlier, at 13:16:59 UTC and 13:17:00 UTC~\citep{2023ApJ...952L..42L, 2023arXiv230301203A}, respectively.

The data used in this paper come from the X-ray telescope (XRT) of the Neil Gehrels {\it Swift} satellite~\citep{2004HEAD....8.1603B}. Working in the photon energy range of 0.3 $-$ 10.0 keV, XRT has two data modes: windowed-timing (WT) mode and photon-counting (PC) mode. The WT mode allows imaging in only one spatial dimension, whereas the PC mode produces two-dimensional images \citep{2004SPIE.5165..217H}. We focused primarily on the PC data, download from the {\it Swift} Science Data Centre\footnote{\url{https://www.swift.ac.uk/user_objects/}}. We analysed 0.3-10 keV images centred on the GRB 221009A location using the {\it XIMAGE}\footnote{\url{https://heasarc.gsfc.nasa.gov/docs/software/lheasoft}} software package. During the processing of the XRT data, we use the {\it xrtpipeline}\footnote{\url{https://swift.gsfc.nasa.gov/analysis/xrt_swguide_v1_2.pdf}} tool to generate the 2-level event files for calibration and cleaning. First we extract the total image using the {\it XSELECT}\footnote{\url{https://heasarc.gsfc.nasa.gov/docs/software/lheasoft/ftools/xselect/xselect.html}} tool, then, we use the annulus in the Ds9\footnote{\url{https://ds9.si.edu/doc/user/gui/index.html}} tool to extract the rings. Finally we extract the corresponding energy spectrum by {\it XSELECT}. The spectral data was binned using the tool {\it GRPPHA}\footnote{\url{https://heasarc.gsfc.nasa.gov/docs/software/lheasoft}} to have a sufficient number of counts in each bin. We used the {\it XSPEC}\footnote{\url{https://heasarc.gsfc.nasa.gov/xanadu/xspec/manual/manual.html}} tool for the X-ray spectra fitting.

We use six XRT observations performed between MJD 59862.34--59866.03 with OBS-ID numbers 01126853004, 01126853005, 01126853006, 01126853008, 01126853009 and 01126853011, respectively. We adopt the {\it Insight}-HXMT trigger time as the burst onset time, i.e., T$_0$= 13:17:00 UTC on 2022 October 9. The start times of the six observations are T$_0$ + 67693 s, 105647 s, 111954 s, 215928 s, 300355 s and 386300 s, respectively. 

\begin{figure*}
    \centering
    \subfigure[]{
        \includegraphics[width=3.4in]{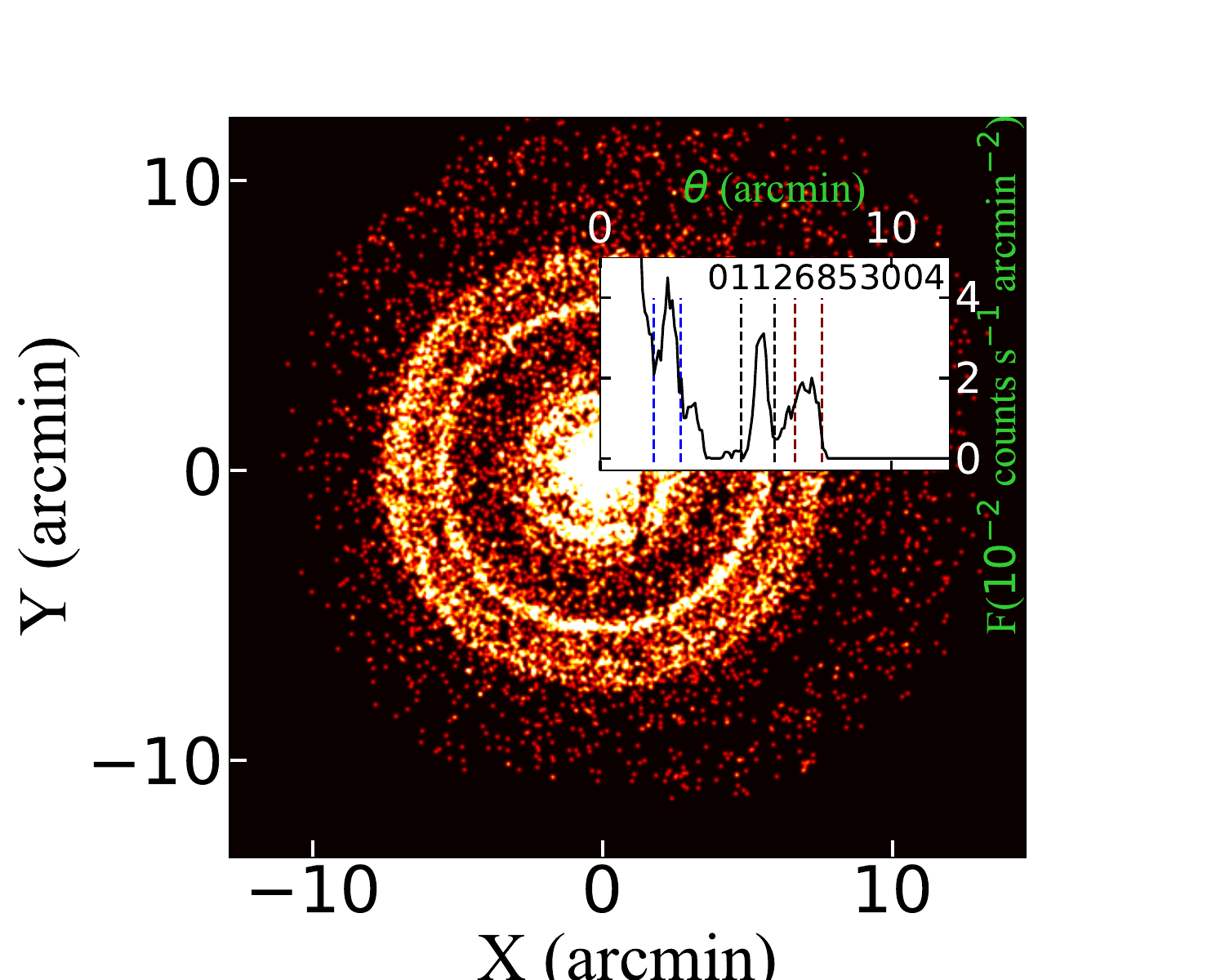}
    }
    \subfigure[]{
	\includegraphics[width=3.4in]{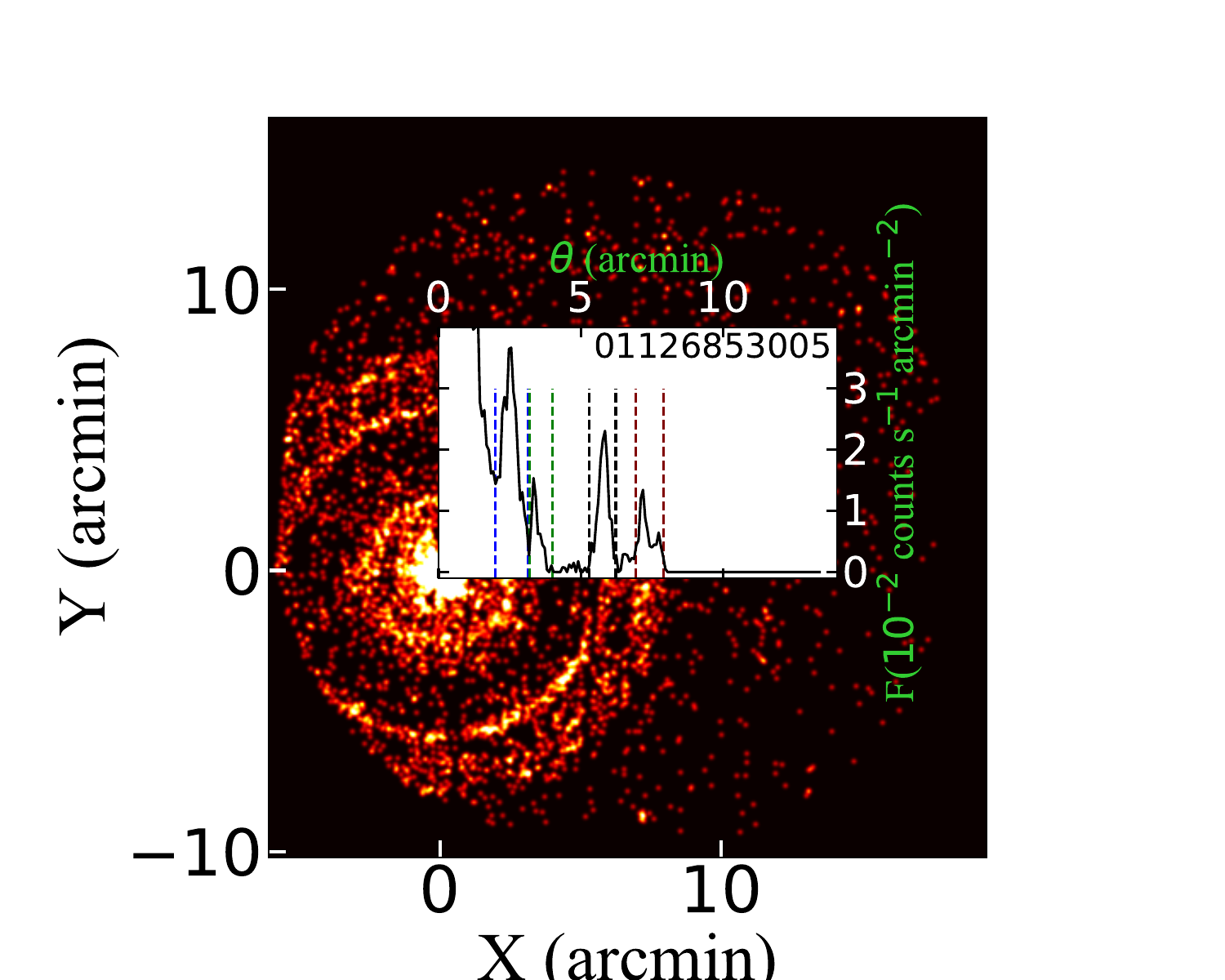}
    }
    \subfigure[]{
    	\includegraphics[width=3.4in]{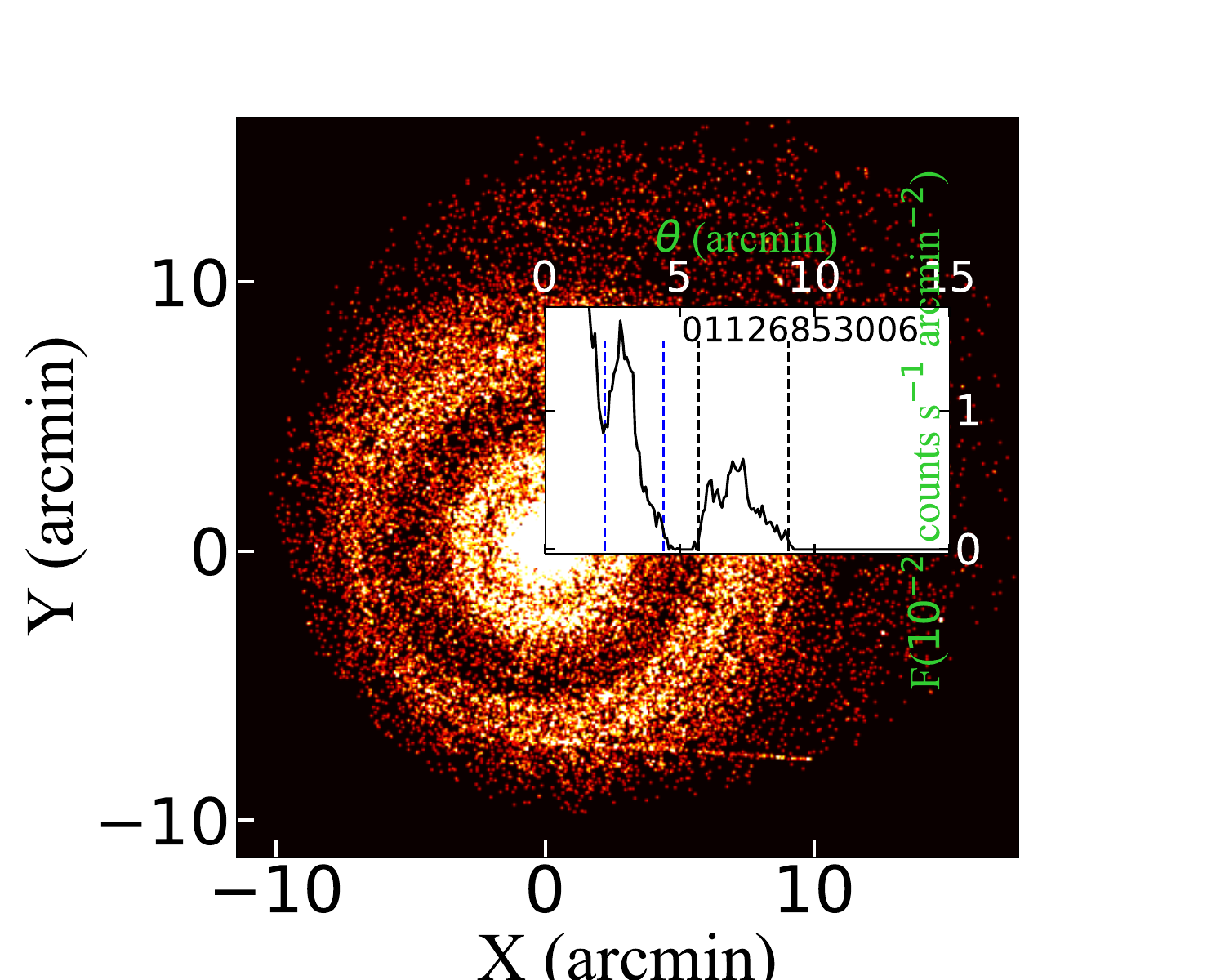}
    }
    \subfigure[]{
	\includegraphics[width=3.4in]{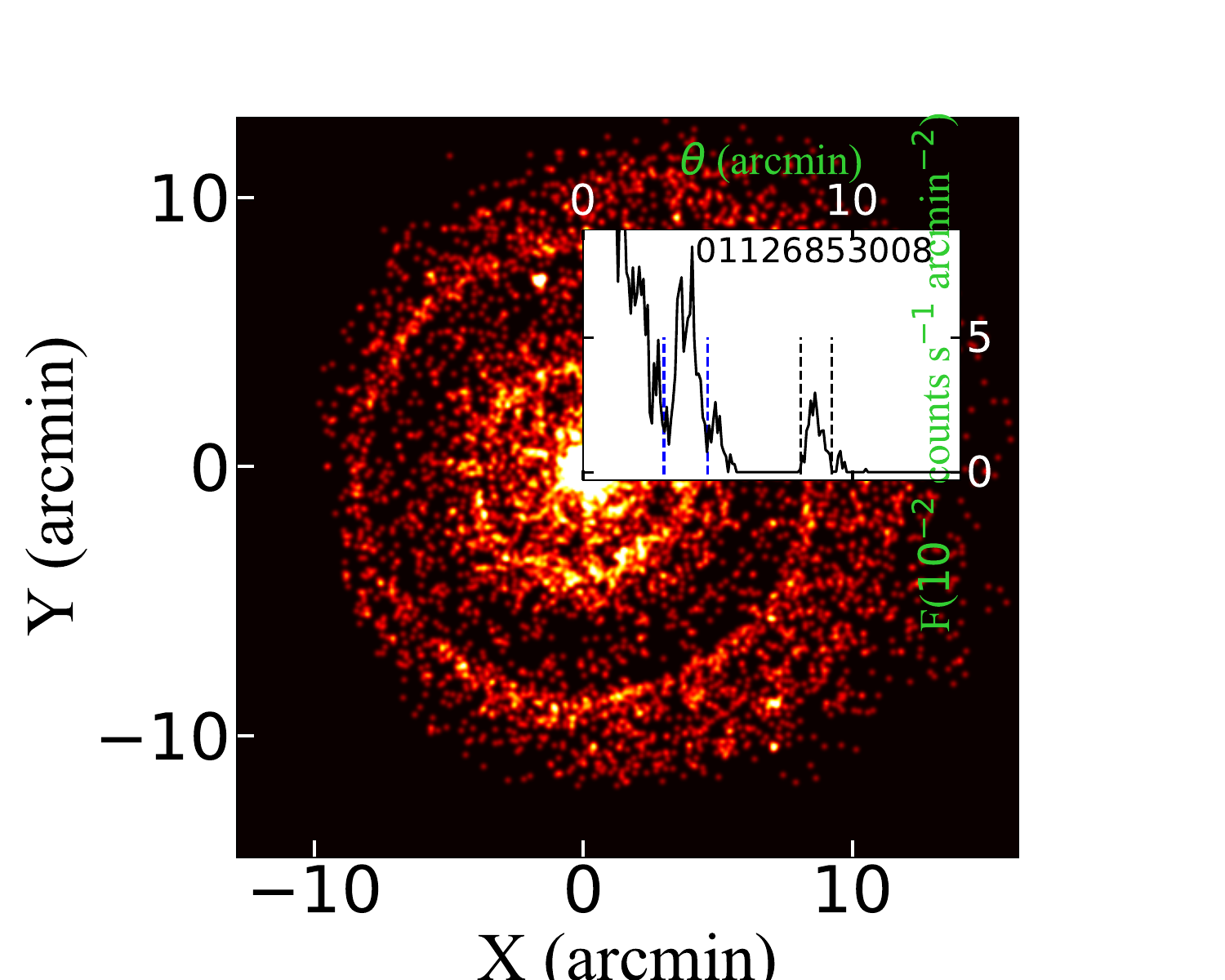}
    }
    \subfigure[]{
    	\includegraphics[width=3.4in]{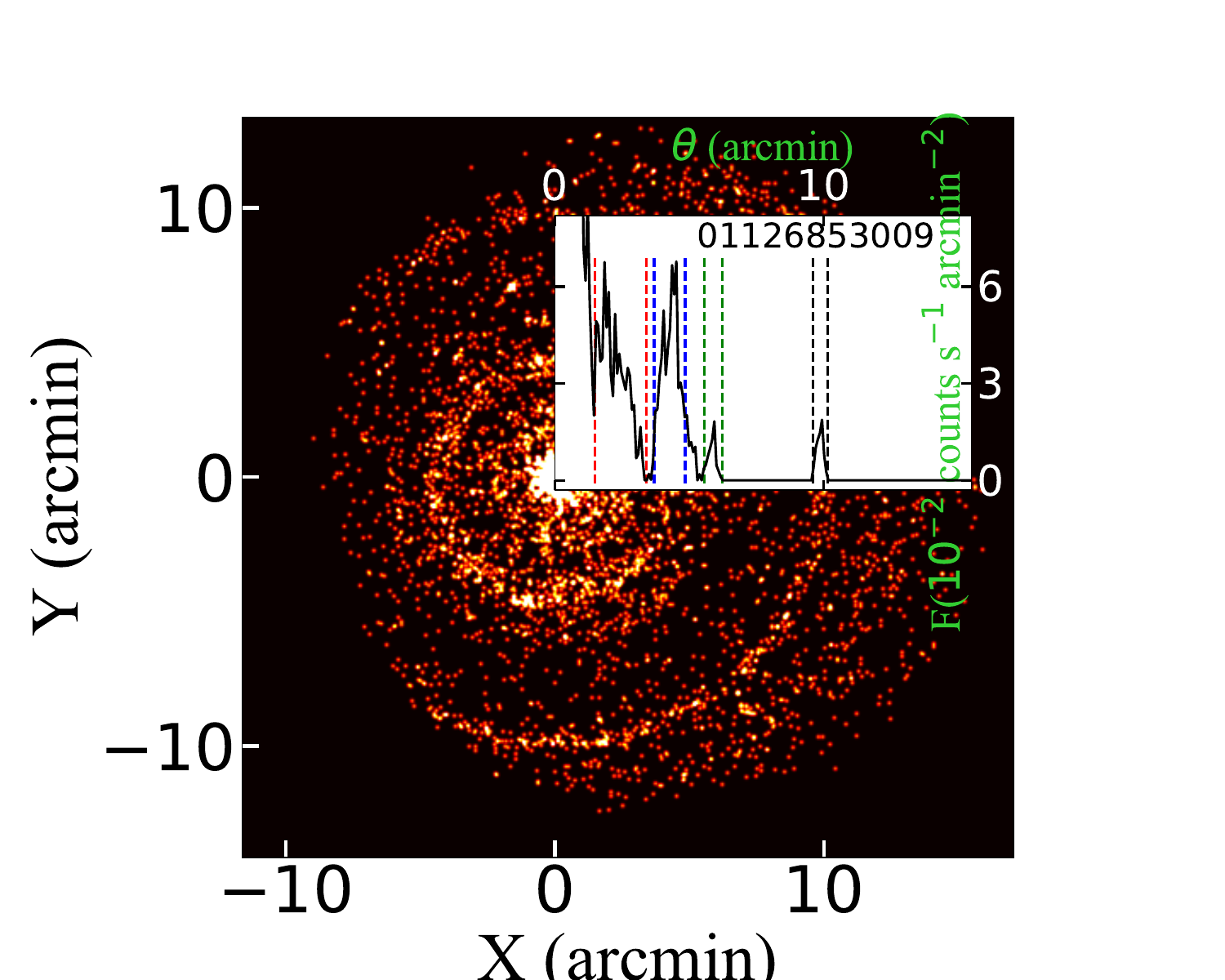}
    }
    \subfigure[]{
	\includegraphics[width=3.4in]{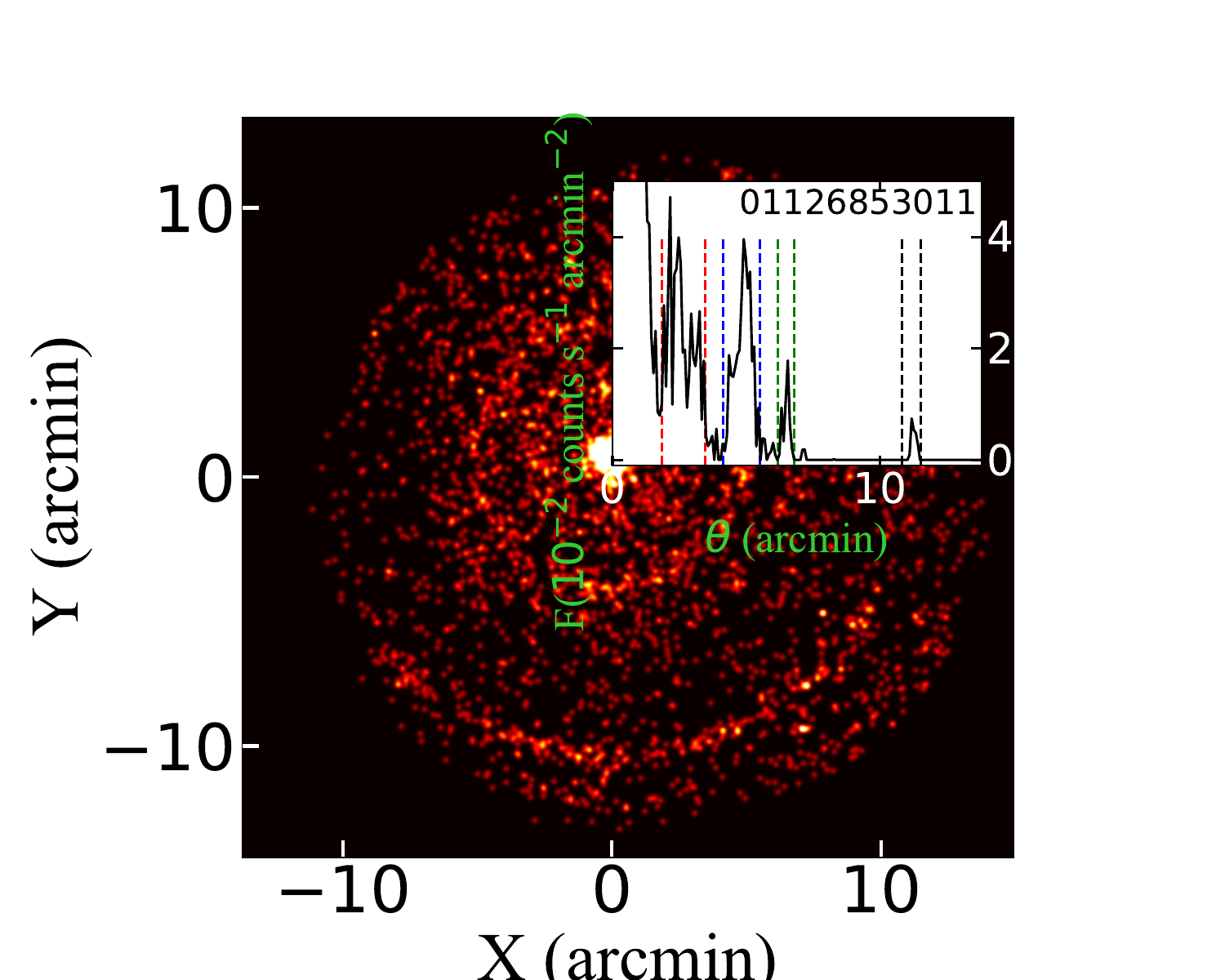}
    }
    
    \caption{{\it Swift}/XRT 0.3-10 keV PC-mode images and the flux radial profile (insets) from six observations. Each pair of vertical dashed lines of the same colour mark the width of the ring.}
    \label{fig:image_psf}
\end{figure*}

We extracted images of these six observations and they are shown in Figure~\ref{fig:image_psf}. All the images show multiple rings and the surface brightness of each ring is not the same. The presence of multiple rings are due to the fact that photons with the same arrival time but with different scattering angles propagate in different paths. We can see that there are at least two rings per observation. To help identify different rings, we plot the radial profile of the X-ray surface brightness of each image in the insets of Figure~\ref{fig:image_psf}. This is obtained by summing the counts in each of a series of continuously distributed concentric annuli (with equal binsize of 1 arcsec) centered on the GRB location. 

Except for the central bright peak which represents the on-going GRB X-ray afterglow, each peak in a single inset of Figure~\ref{fig:image_psf} corresponds to a dust-scattering ring in the image, caused by an individual dust cloud on the line of sight. In total we get nineteen rings from these six observations, each observation having 2-4 rings.

\section{RING EXPANSION AND MODELING} \label{sec:expanding}
 
\begin{figure}
\includegraphics[width=3.5in]{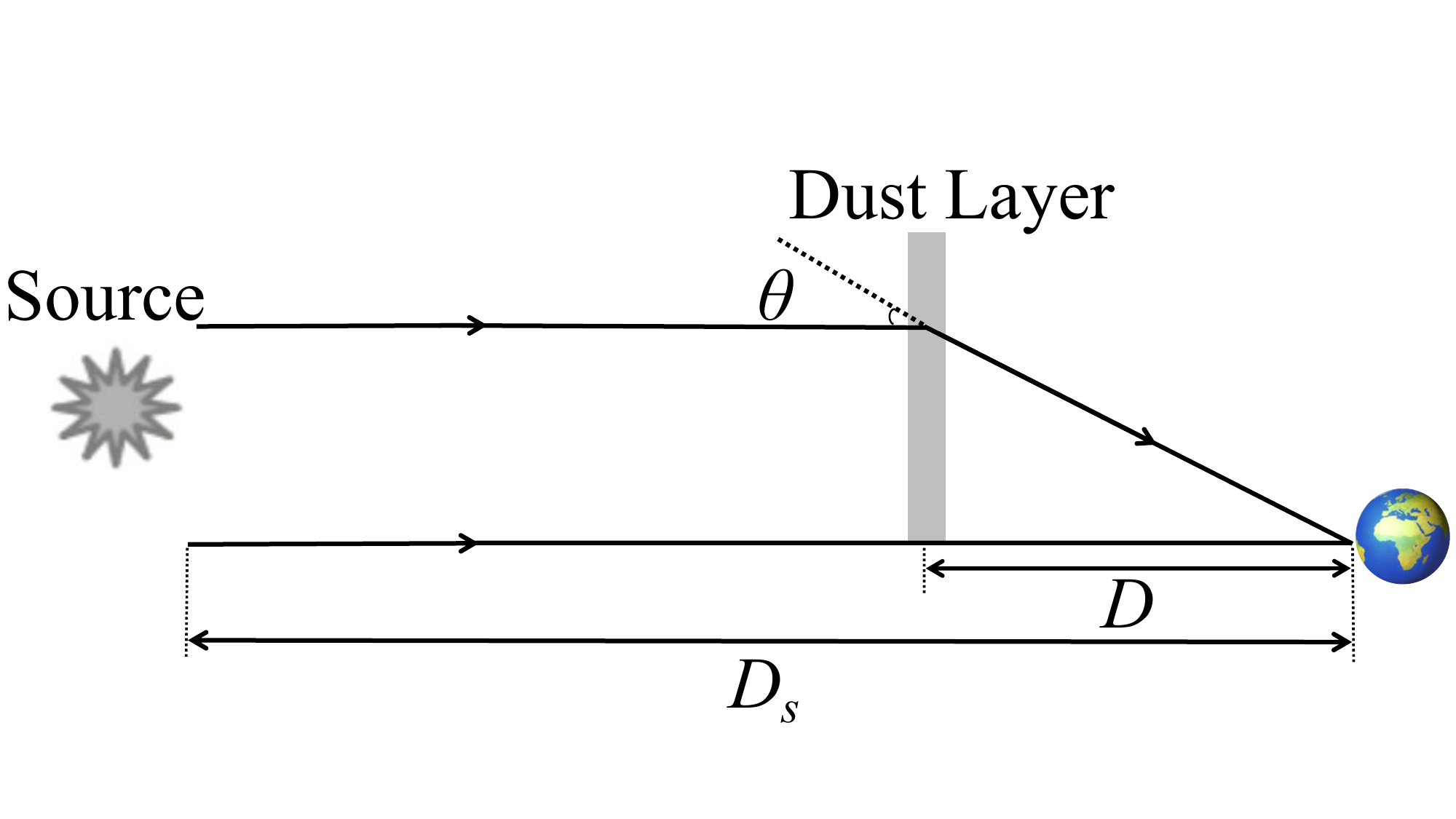}
\caption{Schematic illustration of X-ray scattering in the dust layer at a distance $D$ from the observer. We consider the source at a distance $D_s \gg D$ from the Earth. The scattered photons at a scattering angle $\theta$ will be observed with a time delay {\it $t$}, with respect to the photons that reach the observer without scattering. }
\label{fig:dust}
\end{figure}

The dust scattering echoes contain a wealth of information about the Galactic dust distribution along the line of sight of GRB 221009A, providing a unique opportunity to measure distances to those dust clouds. The geometry of X-ray scattering by a single dust cloud is shown in Figure~\ref{fig:dust}. We consider the source at a distance $D_s$ from the earth, whose X-ray photons can be scattered at a small angle $\theta$ by the dust layer at a distance $D$, and the scattered photons will be observed with a time delay {\it $t$}. 

The distance to the scattering dust can be inferred from the radial expansion of the ring as its size increases with time. For a ring detected at a delay time {\it $t$}, the scattering angle $\theta(t)$ (also the ring angular size) is given by ~\citep{2006A&A...449..203T}
\setlength\abovedisplayskip{3pt}
\setlength\belowdisplayskip{3pt}
\begin{equation}
\theta (t) \approx \left(\frac{2ct}{D}\right)^{1/2} = 4.8\left(\frac{1 \mathrm{kpc}}{D}\right)^{1/2}\left(\frac{t}{10^{5}\mathrm{s}}\right)^{1/2}~\mathrm{arcmin}, \label{eq:sacttering_distance}
\end{equation}
where $D$ is the distance between the dust and the observer, {\it c} is the speed of light, and $t=0$ is the arrival time of the photon at zero scattering angle, and also the arrival time of the GRB prompt emission. Photons scattered by the same dust layer but arriving later will constitute a larger ring. Thus, the observer sees an expanding ring.

\begin{figure}
\includegraphics[width=3.6in]{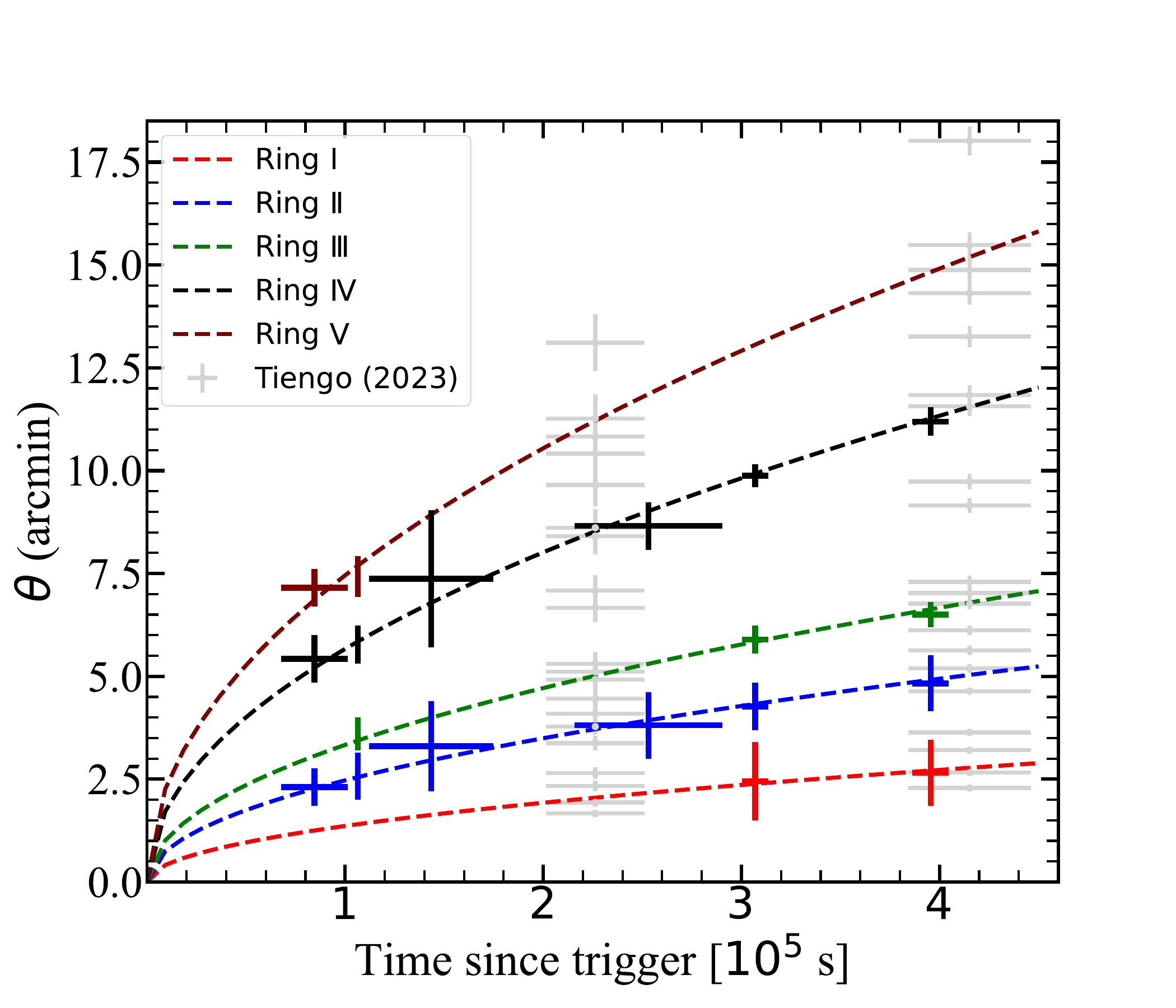}
\caption{Evolution of the angular size of the detected rings (crosses) in the {\it Swift}/XRT images. The error bars of the radii represent the widths of the rings. The ring expansion is well fitted by equation~\eqref{eq:sacttering_distance} (dashed lines). For reference, the grey crosses are 20 expanding rings obtained by \protect\cite{2023ApJ...946L..30T} from two XMM-Newton observations.} Each colour represents the same possible expanding ring.
\label{fig:ring}
\end{figure}

We plot the angular radii of the rings identified in Section \ref{sec:obs}, vesus the observing times, in Figure~\ref{fig:ring}. They are then fitted with Equation~\eqref{eq:sacttering_distance}~(dashed lines), which describes the radial expansion of the ring with time. The data and the fit in Figure~\ref{fig:ring} show that five expanding rings can be identified, each being indicated by the same color and corresponding to one dust cloud. The best-fit model in Figure~\ref{fig:ring} give the distances of the five dust clouds to be 0.42 $\pm$ 0.03~kpc, 0.72 $\pm$ 0.02~kpc, 2.1 $\pm$ 0.07~kpc, 3.8 $ \pm $ 0.1~kpc and 12.4 $\pm$ 0.6~kpc, respectively.

\section{SPECTRAL EVOLUTION AND MODELING} \label{sec:SPECTRAL}
The spectral information of the ring flux may carry the characteristics of the dust scattering process. We extracted the PC-mode spectrum of each ring using an annulus that encompasses the ring, and extracted the darker signals in the region outside it as the background spectrum. We substract the background spectrum from the spectrum of interest to reduce the interference of noise. Then we fit it with an absorbed power-law model $\mathrm{N(E)}\propto E^{-\Gamma}$ where $\Gamma$ is the photon index; the other two parameters are a normalisation parameter and an equivalent Glactic hydrogen column density $\mathrm{N_{H}}$. The fitting was done using Xspec version 12.12.1 \citep{1996ASPC..101...17A} utilizing the absorption model TBABS \citep{2000ApJ...542..914W}.

\begin{figure}
\includegraphics[width=3.6in]{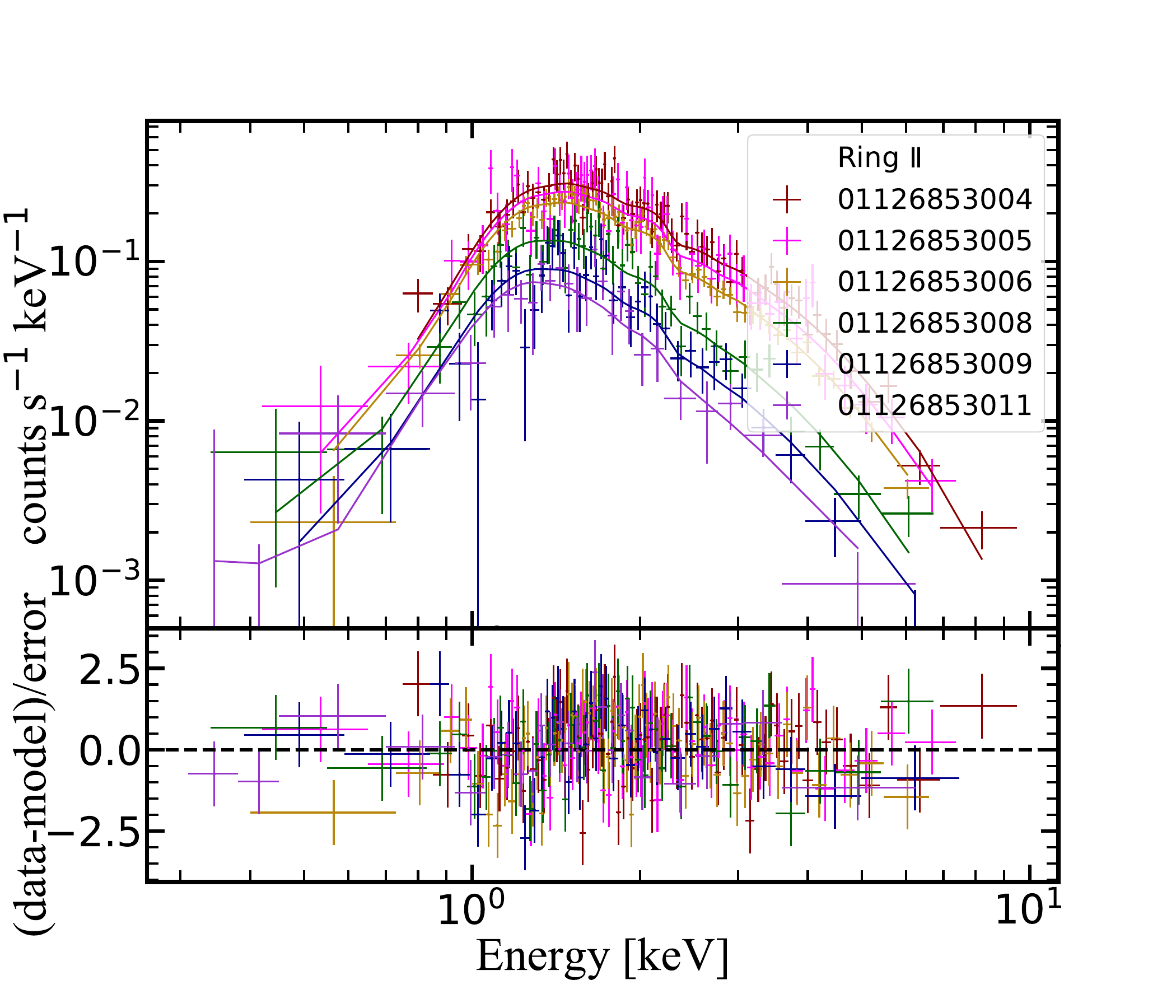}
\caption{The 0.3-10 keV spectra of ring \uppercase\expandafter{\romannumeral2} for multiple exposure times and the best fitting models. Top panel: spectrum is fitted with the power-law model. The solid lines of different colours represent the best models for different exposure times.  Bottom panel: the correlation residuals of the best-fit model for each spectrum.}
\label{fig:spec_fit}
\end{figure}

\begin{figure}
\includegraphics[width=3.6in]{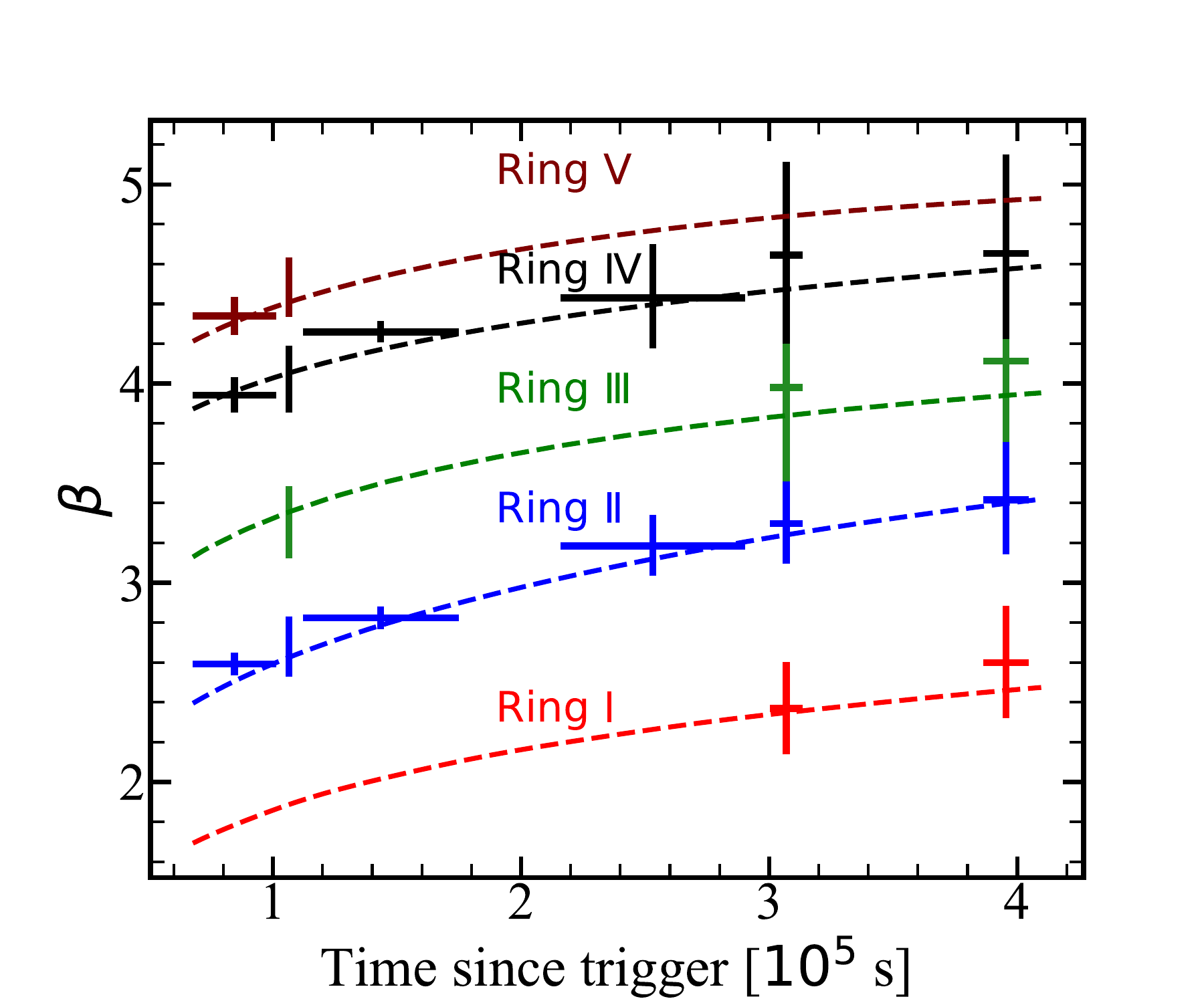}
\caption{The spectral index as a function of time for the five expanding rings. The spectral softening for each expanding ring is clearly seen. The dashed lines are the model-predicted evolution of the spectral index with time, computed with Eq.~\eqref{eq:index} using the dust clouds distances obtained in Section~\ref{sec:expanding} and Figure~\ref{fig:ring}. The parameter values for the dust grain size distribution ($a_-$ and $a_+$ and $q$) are adjusted so as to get the best fit.}
\label{fig:spec_model}
\end{figure}

As an example, we show the spectral fit results for the expanding ring \uppercase\expandafter{\romannumeral2} in Figure~\ref{fig:spec_fit}. It shows that as time increases, the spectral peak gradually moves to lower energies, indicating a gradual softening of the spectrum. The spectral fit results for all expanding rings show a range of the spectral index, of $\beta$ = 2.2$\sim$5, with a gradual softening, i.e., an ever increasing $\beta$, for each expanding ring. Here the spectral index is defined as $F(E)\propto E^{-\beta}$ and $\beta = \Gamma -1$. The best fit absorbing H-column density is $N_\mathrm{H}$ = 1.7$\times$ 10$^{22}$~cm$^{-2}$. The evolution of the spectral index over time is shown in Figure~\ref{fig:spec_model} (as crosses). Each same color in Figure~\ref{fig:spec_model} corresponds to the same expanding ring in Figure~\ref{fig:ring}.

This spectral softening is consistent with the predictions of the dust scattering model of \cite{2007ApJ...660.1319S} and \cite{2009MNRAS.393..598S}. Below, we briefly review this model. We assume that the grains in the dust cloud (here simplified to a dust layer as shown in Figure~\ref{fig:dust}) have a power-law size distribution within a size range ($a_-, a_+$), i.e., the column density $[$cm$^{-2}]$ of the grains whose sizes are in the interval $(a, a+da)$ is given as \citep{1977ApJ...217..425M, 1986ApJ...302..371M, 2003ApJ...598.1026D}
\setlength\abovedisplayskip{3pt}
\setlength\belowdisplayskip{3pt}
\begin{equation}		
\label{eq:dna}
d N(a) = N_{0.1 \mu m} \left(\frac{a}{0.1 \, \mu \mbox{m}}\right)^{-q} da,
\end{equation}
where $N_{0.1~\mathrm{\mu m}} \, [\mbox{cm}^{-2}~\mu\mbox{m}^{-1}]$ is the column density per unit size interval at $a=$ 0.1 $\mu$m, and $q$ is the distribution index. 

For a photon of energy $E$ (or wavelength $\lambda$) to be scattered at small angles by a dust grain of size $a$, the differential cross section of the scattering $[$cm$^2$ strd$^{-1}]$ in the Rayleigh-Gans approximation is \citep{1965ApJ...141..864O, 1978ApJ...222..456A, 1986ApJ...302..371M} \footnote{Note that in Eq.(8) of \cite{2007ApJ...660.1319S} a factor of $a/\lambda$ is missing.}
\begin{equation}
\label{eq:dif-sigma}
\begin{aligned}
&\frac{d\sigma (E, a, \theta)}{d\Omega} = 8\pi \left(\frac{a}{\lambda}\right)^2 \sigma_t(E,a)  \left[\frac{j_1(x)}{x}\right]^2,
\end{aligned}
\end{equation}
where $\sigma_t(E,a)\approx 6.3\times 10^{-11}\left(\frac{E}{1 \mathrm{keV}}\right)^{-2}\left(\frac{a}{0.1 \mathrm{\mu m}}\right)^{4}~\mathrm{cm^2}$ is the total scattering cross section. In the angular-dependence part on the right hand side of Eq.~\eqref{eq:dif-sigma}, $j_1(x) = \frac{(\sin x - x \cos x)}{x^2}$ is the first-order spherical Bessel function, and 
\begin{equation}
\label{eq:dif-sigma}
\begin{aligned}
x \equiv \frac{2\pi a}{\lambda}  \theta = \frac{2\pi a}{\lambda}  \sqrt{\frac{2ct}{D}}
\end{aligned}
\end{equation}
is the scaled scattering angle. 

Consider that a GRB source, during its prompt emission phase, shines on the dust screen an X-ray energy per unit area $[$erg cm$^{-2}]$ of
\begin{equation}
d \mathcal{F}(E) = S(E) dE,
\end{equation}
within the photon energy interval $(E, E+dE)$, where $S(E)$ $[$erg cm$^{-2}$ keV$^{-1}]$ is the photon fluence per unit photon energy interval. Since the GRB prompt phase lasts very briefly ($\sim$ 10 s) compared with the dust echo, it can be considered as being instantaneous.

Consider an annulus of $(\theta, \theta+ d\theta)$ on the dust layer that was shown in Figure~\ref{fig:dust}. The solid angle opened by the annulus with respect to the observer is $d \omega(\theta)= 2\pi \theta d \theta$. The radiation energy per unit surface area on the annulus that is scattered toward the observer within a unit solid angle $[$erg~cm$^{-2}$~strd$^{-1}]$ would be 
\setlength\abovedisplayskip{3pt}
\setlength\belowdisplayskip{3pt}
\begin{equation}
d \mathcal{F}(E) d N(a) \times \frac{d \sigma (E, a, \theta)}{d\Omega}.
\end{equation}
All the scattered photons from this annulus will arrive at the observer within the time interval $(t, t+dt)$, and $dt= (D/c)\theta d\theta$. Then the flux $[$erg~s$^{-1}$~cm$^{-2}]$ received at the observer from the annulus is 
\setlength\abovedisplayskip{3pt}
\setlength\belowdisplayskip{3pt}
\begin{equation}
\begin{aligned}
d F_{E,a}(\theta) &= \frac{d \mathcal{F}(E) d N(a)}{dt} \times \frac{d \sigma (E, a, \theta)}{d\Omega}~d \omega (\theta)\\& =\frac{2 \pi c}{D} S(E) dE~d\tau(E,a,\theta),
\end{aligned} 
\end{equation}
where $d \tau(E, a, \theta) \equiv  d N(a) \times d \sigma / d \Omega$ is the differential optical depth of the dust layer for scattering.

Terefore, at time $t$, the detector received a scattered specific flux $[$erg cm$^{-2}$ s$^{-1}$ keV$^{-1}]$ from the dust grains in the size range $(a_-,a_+)$ at the photon energy $E$,

\begin{equation}
\label{eq:F_E(t)}
\begin{aligned}
F_{\mathrm{E}}(t) &= \int_{a_{-1}^-}^{a_{-1}^+} \frac{dF_{E,a}(\theta)}{dE} = \frac{2}{t} S(E)\int_{a_{-1}^-}^{a_{-1}^+} \sigma_t(E,a) dN(a) j_1^2(x)  \\&= \frac{2 \tau_0}{t} S(E) \left(\frac{E}{\rm 1\, keV}\right)^{-2} \int_{a_{-1}^-}^{a_{-1}^+} j_1^2(x) a_{-1}^{4-q} d a_{-1},
\end{aligned}
\end{equation}

where $a_{-1} \equiv a/({\rm 0.1\, \mu m})$ and the nominal optical depth (dimensionless) of the dust layer is 
\setlength\abovedisplayskip{3pt}
\setlength\belowdisplayskip{3pt}
\begin{equation}
\tau_0 \equiv \sigma_t(E,a)  a  \frac{dN(a)}{da} \Bigg|_{E= 1\, {\rm keV},\, a= 0.1 \, \mu{\rm m}}= 6.3\times 10^{-12}~N_{0.1 \mu{\rm m}}. 
\label{eq:tau}
\end{equation}
Thus, the flux received by an X-ray detector within an energy range ($E_-,E_+$) at time $t$ is
\setlength\abovedisplayskip{3pt}
\setlength\belowdisplayskip{3pt}
\begin{equation}
\begin{aligned}
F_\mathrm{X}(t)=\int_{E_-}^{E_+}F_E(t) dE.
\label{eq:flux-t}
\end{aligned}
\end{equation}

It is easy to show that the angular term $j_1^2(x)$ increases as $\propto x^2$ from $x=0$ to $x \simeq 1.5$ and then drops rapidly as $\propto x^{-2}$ for $x > 1.5$ \citep{1970PThPh..43.1224H}. Therefore, at a given photon energy $E$, the echo ring flux first evolves as a plateau with time ($\propto t^0$), then decays as $\propto t^{-2}$. The transition time would be given by
\begin{equation}
t_c= 4.5\times10^4 \left(\frac{E}{{\rm 1\, keV}}\right)^{-2} \left(\frac{D}{100\, {\rm pc}}\right) \biggl(\frac{a}{0.1\, {\rm \mu m}}\biggr)^{-2}\,\, {\rm s},
\end{equation}
which corresponds to a characteristic scattering angle $x \simeq 1.5$,   or 
\begin{equation}
\theta_c= 10 \left(\frac{E}{{\rm 1\, keV}}\right)^{-1} \left(\frac{a}{\rm 0.1\, \mu m} \right)^{-1} \, \mbox{arcmin}.
\end{equation}
The photon energy dependences of $\theta_c$ and $t_c$ suggest that the decay of the ring flux starts much earlier at higher photon energy than at lower energies. Thus the overall echo emission must experience strong spectral softening.

To compare with the spectral fit result from the data, we compute a model-predicted two-point spectral index as
\begin{equation}
\label{eq:index}
\beta(\mathrm{t}) = \mathrm{log}\left(\frac{F_\mathrm{E=0.3 keV}}{F_\mathrm{E=10 keV}}\right) ~/~\mathrm{log}\left(\frac{10}{0.3}\right),
\end{equation}
where F$_\mathrm{E}$(t) is calculated by Eq.~\eqref{eq:F_E(t)}, and the dust cloud distance $\mathrm{D}$ takes the values obtained from the ring expansion modeling (Section \ref{sec:expanding}).

The GRB prompt X-ray fluence S(E) is needed as well to calculate F$_\mathrm{E}$(t), which we obtain as follows. We first fit the Band function~\citep{1993ApJ...413..281B} to the prompt {\it Fermi}/GBM $\gamma$-ray spectral data\footnote{\url{https://heasarc.gsfc.nasa.gov/FTP/fermi/data/gbm/}}~\citep{2023ApJ...952L..42L}. The best-fit parameters are listed in Table~\ref{table:GBM}. Due to the superior fitting results of the detector n6 data, we choose to adopt its best-fit parameter values: the lower-energy photon index $\alpha_\mathrm{L} = -1.48$, the high-energy photon index $\alpha_\mathrm{H} = -1.68$, E$_\mathrm{peak}$ = 120 keV and the normalization $A = \mathrm{3.29\times 10^{-4}~counts~cm^{-2}~keV^{-1}}$.  
Then we extrapolate the fitted Band function to the X-ray band, and get $\mathrm{S(E=0.3~keV) = 1.6\times 10^{-3}~erg~cm^{-2}~keV^{-1}}$and $\mathrm{S(E=10~keV) = 9.5\times 10^{-3}~erg~cm^{-2}~keV^{-1}}$. Note that in a different approach, \cite{2023ApJ...946L..30T} assumed different dust compositions and grain size distribution models to fit the spectrum of the ring, and reconstructed GRB 221009A's prompt X-ray spectrum. They obtained $\mathrm{S(E) = 3\times 10^{-4} \sim 2 \times 10^{-3}~erg~cm^{-2}~keV^{-1}}$ in the 0.5$-$5 keV range.

\begin{table}
\caption{Best-fit spectral parameters of Band function to the prompt Fermi GBM data of GRB 221009A.}\label{table:GBM}
\centering
\begin{threeparttable}          
    \begin{tabular}{*5{c}}
    \hline 
    Detector & $\alpha_{\mathrm{L}}$ & $\alpha_\mathrm{H}$  & $\mathrm{E_{peak}~[keV]}$ &C$_\mathrm{stat}/D.o.F$\\ 
    \hline 

    n3& -0.85& -2.2  & 100 &328/117\\
    n4& -1.38 &-1.51  & 56 &298/115\\
    n6& -1.48 & -1.68  & 120 &293/115\\
    n7& -1.53 & -1.79  & 98 &328/115\\
    n8& -1.38 & -1.69  & 113 &491/117\\
    \hline 

    \end{tabular}
         
\end{threeparttable}  
\end{table}

We fit the observed spectral index evolution of each ring shown in Figure \ref{fig:spec_model} with Eq.~\eqref{eq:index}. In this way we also determine the size of the dust grains ($a_-$ and $a_+$) as well as the exponent {\it q}. The best-fit parameters are shown in Table~\ref{table:fit}. The calculated spectral indices for each ring are plotted as the dashed lines in Figure~\ref{fig:spec_model}, which well reproduce the observed spectral evolution. Altogether, this consistency suggests that the spectral softening is a robust feature from the X-ray dust scattering model.

\section{RING FLUX EVOLUTION AND DUST COLUMN DENSITY} \label{sec:density}
\begin{table*}
\caption{Best-fit parameters for dust scattering rings.}\label{table:fit}
\centering
\begin{threeparttable}          
    \begin{tabular}{*8{c}}
    \hline 

    Ring & $D[\mathrm{kpc}]^{\mathrm{a}}$& $D_v[\mathrm{kpc}]^{\mathrm{b}}$ & $a_-[\mathrm{\mu m}]$  & $a_+[\mathrm{\mu m}]$ &$q$ &$N_{0.1\mu m}[ 10^{7}~\mathrm{cm^{-2}\mu m}^{-1}]^{\mathrm{c}}$ & $N[ 10^{8}~\mathrm{cm}^{-2}]^{\mathrm{d}}$\\ 
    \hline 

    \uppercase\expandafter{\romannumeral1} & $12\pm 0.6$ &  & 0.015  & 0.55 & 3.8& 0.9& 0.7\\
    \uppercase\expandafter{\romannumeral2} & $3.8\pm 0.1$ & $3.7\pm 0.06$& 0.015 & 0.65 &3.7 &1.1 & 0.7\\
    \uppercase\expandafter{\romannumeral3} & $2.1\pm 0.07$ & $2.0\pm 0.03$& 0.015  & 0.75 &3.0 &0.7 & 0.5\\
    \uppercase\expandafter{\romannumeral4} & $0.72\pm 0.02$&$0.69\pm 0.01$ & 0.015  & 0.60 &4.0 &4 & 4.0 \\
    \uppercase\expandafter{\romannumeral5} & $0.42\pm 0.03$  & $0.44\pm 0.01$& 0.015  & 0.55 &3.6&2.1 & 1.1 \\
    \hline 

    \end{tabular}
        \begin{tablenotes}  
    \footnotesize              
    \item[a] Distance from the dust cloud to Earth. ${\mathrm{^{b}}}$ Dust cloud distances obtained by~\cite{2023MNRAS.521.1590V}. ${\mathrm{^{c}}}$ The column density per unit size interval at $a = 0.1 \mu$m of the cloud. ${\mathrm{^{d}}}$ The column density of all grains for the cloud.
    \end{tablenotes}            
\end{threeparttable}  
\end{table*}

\begin{figure}
\centering
\includegraphics[width=3.6in]{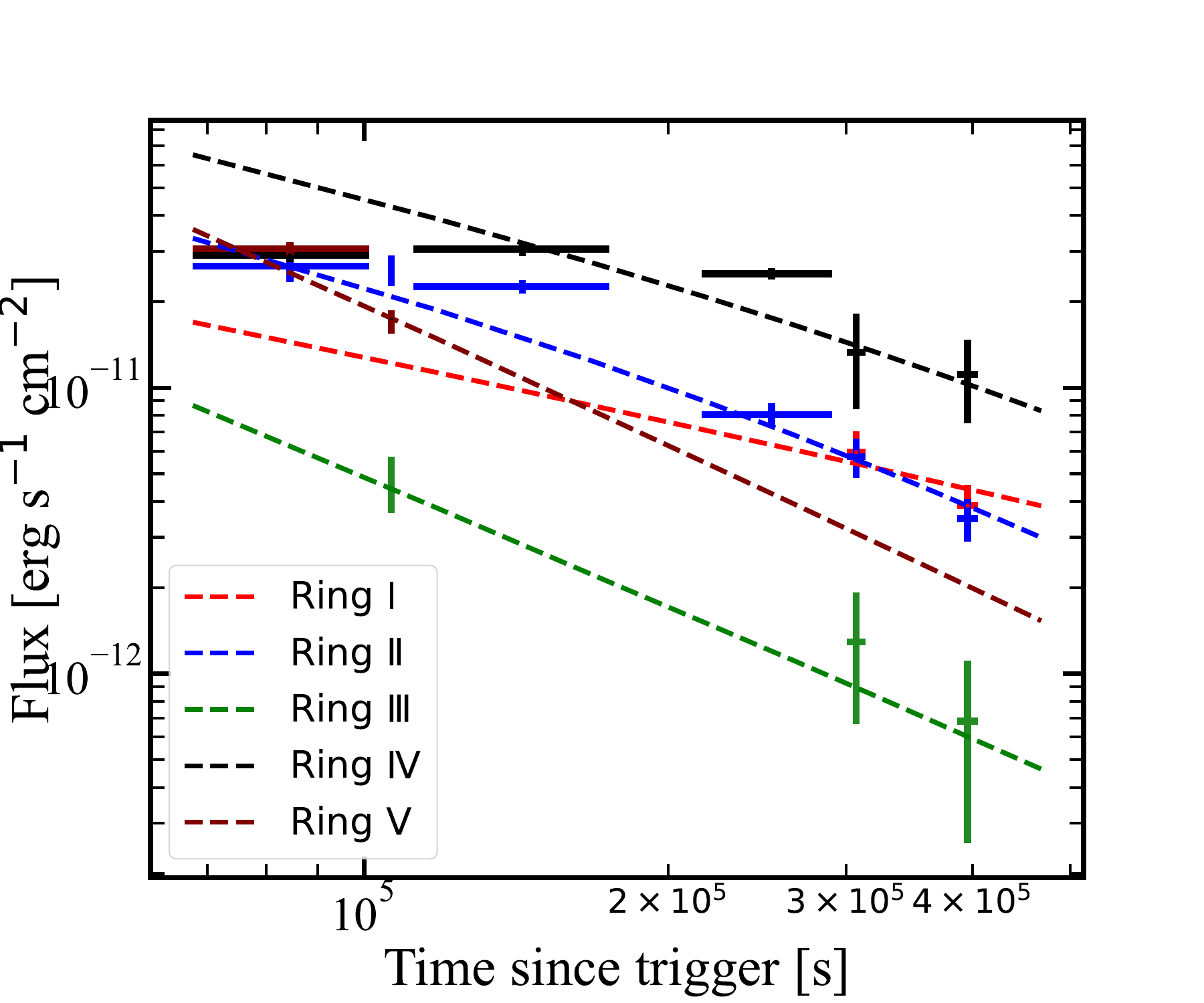}
\caption{Flux evolution of dust scattering rings. The crosses represent the fluxes obtained from the observations, while the dashed lines represent the fluxes of the dust rings obtained from Eq.~\eqref{eq:flux-t}, with different colors representing different expanding dust rings.}
\label{fig:flux_tau}
\end{figure}

Note that Eqs.~(\ref{eq:F_E(t)}-\ref{eq:flux-t}) not only predict the decline of the ring flux, and but also contain the information about the dust column density of the cloud (included in $\tau_0$). Here we compare them with the data and infer the column density. We fitted the observed spectrum using a simple power-law function to obtain the flux [$\mathrm{erg~s^{-1}~cm^{-2}}$] of each ring, plotted as crosses in Figure~\ref{fig:flux_tau}.

The model-predicted flux evolution of the expanding ring was computed using Eq.~\eqref{eq:flux-t}, with $E_-$ = 0.3 keV and $E_+$ = 10 keV, and using the best-fit parameter values of $a_-$, $a_+$ and $q$ obtained in Section~\ref{sec:SPECTRAL} and listed in Table~\ref{table:fit}; S(E) takes the value at 0.3-10 keV extropolated from the Band function fitting (Section ~\ref{sec:SPECTRAL}). It is shown as the dashed lines in Figure~\ref{fig:flux_tau}.  

Our analysis leads to an estimate of N$_{0.1 \mathrm{\mu m}} $ in the range of $(0.7-4)\times10^{7}$~cm$^{-2}~\mu$m$^{-1}$. The column density of dust particles of all sizes in the dust cloud is estimated by integrating Eq.~\eqref{eq:dna} over ($a_-,a_+$), which gives $\mathrm{N = (0.5-4)\times 10^{8}~cm^{-2}}$. The best-fitting parameters are provided in Table~\ref{table:fit}. Note that \cite{1986ApJ...302..371M} and \cite{2010ApJ...717..223H} estimated $\mathrm{N\approx 10^{9}~cm^{-2}}$ and 10$^{12}$~cm$^{-2}$, respectively, using the scattering optical depth. Taking a typical size of cloud $\mathrm{R\approx10~pc}$, the number density of grains in the dust cloud is estimated to be $\mathrm{n\approx N/R\sim 10^{-12} - 10^{-11}~cm^{-3}}$.
\section{Comparison with previous works} \label{sec:Comparison}

Form nineteen rings seen in XRT observations, we identified five expanding rings. Previous studies have already analyzed scattering rings of GRB 221009A~\citep{2023ApJ...946L..21N,2023ApJ...946L..24W, 2023ApJ...946L..30T,2023MNRAS.521.1590V}. Here, we discuss the similarities and differences between our findings and theirs. 

\cite{2023ApJ...946L..21N} obtained two rings from two observations with IXPE, while~\cite{2023ApJ...946L..24W} reported a series of dust rings scattered by dust clouds located from 700 pc to over 10 kpc. However, they did not provide the physical connections between these rings, i.e., whether some of them belong to the same ring in expansion, thus preventing direct comparison of our results with theirs.

~\cite{2023ApJ...946L..30T} identified 20 expanding rings seen by XMM (shown as light grey crosses in Figure~\ref{fig:ring}). The discrepancies in results between ours and theirs could be due to the use of different telescope data and methods for identifying the rings. Our ring identification was done by visual inspection, which tends to pick the most prominent ones only. This may have led to the omission of some weaker rings, such as those small peaks in Figure ~\ref{fig:image_psf}. Therefore it is likely that one of the expanding rings we observed may encompass several neighboring expanding rings in their results.

\cite{2023MNRAS.521.1590V} identified five prominent expanding rings from five XRT observations. The rings they reported, as $\mathrm{\uppercase\expandafter{\romannumeral1}}$ to $\mathrm{\uppercase\expandafter{\romannumeral4}}$, correspond to our rings $\mathrm{\uppercase\expandafter{\romannumeral2}}$ to $\mathrm{\uppercase\expandafter{\romannumeral5}}$, respectively (see Table ~\ref{table:fit}). It is worth noting that our ring $\mathrm{\uppercase\expandafter{\romannumeral1}}$, identified from two later observations, was not included in their findings. This is because their analysis used one of the later observations but not the latest one (ObsID 01126853011), and it needs at least two observations to confirm an expanding ring. We did not observe ring $\mathrm{\uppercase\expandafter{\romannumeral5}}$ mentioned in their results, possibly because our analysis employed a relatively simpler method, leading to the omission of some less prominent rings. Another reason for the differences could be that their analysis focused on photons with energies $\geq$ 1 keV, while our analysis included all photons within 0.3-10 keV, resulting in a different total photon counts collected.

Overall, the number of detected expanding rings varied among studies, with some studies identifying more rings than others. This variation could be due to differences in observation telescopes, data processing methods, and energy ranges.

Furthermore, our analysis primarily focuses on the spectral softening of rings (Section~\ref{sec:SPECTRAL}) and the dust column density (Section~\ref{sec:density}), aspects that are not extensively discussed in those studies.


\section{conclusions} \label{sec:con}
In this study, we performed a comprehensive analysis of the X-ray dust scattering rings associated with GRB 2210019A, the most luminous GRB ever recorded. Typically, the scattered radiation from a bright X-ray point source manifests as an X-ray halo. In the particular case of variable X-ray sources that exhibit short bursts, along with non-uniform distribution of dust along the line of sight that concentrates within discrete dust clouds, multiple X-ray rings can be observed. 

Using publicly available data from the {\it Swift}/XRT, we conducted an analysis of six observations of GRB 221009A. The XRT images unveiled the presence of multiple dust halos, from which we identified nineteen dust scattering rings via radial profiles of flux. By employing a simple yet well-established theoretical framework for the scattering of X-ray photons by dust clouds along the line of sight, we identified five expanding rings and successfully modeled their expansion. Our findings indicate that these scattering rings originate from dust clouds situated at distances of 0.42 $\pm$ 0.03~kpc, 0.72 $\pm$ 0.02~kpc, 2.1 $\pm$ 0.07~kpc, 3.8 $ \pm $ 0.1~kpc and 12 $\pm$ 0.6~kpc from the observer.

One notable feature of the dust scattered emission is the spectral softening in the ring flux~\citep{2007ApJ...660.1319S, 2008ApJ...675..507S, 2009MNRAS.393..598S, 2014ApJ...789...74Z}, which derives from the scattering cross section's dependence on photon wavelength~\egcite{1965ApJ...141..864O}{~\ref{sec:SPECTRAL}}. However this prediction has not yet been tested against observational data. Our analysis revealed a significant spectral softening of the ring emission as the rings expand in size, which we accurately reproduce via the existing theoretical framework for the dust scattering of the prompt X-ray emission.

More observations of dust scattering rings may hold the potential to study in detail the dust distribution and composition within our galaxy.

\section*{ }  
We thank Long Ji and Bin-Bin Zhang for their helpful discussion during the early stage of this work. This work is supported by National Natural Science Foundation of China (grants 12261141691 and 12073091).
\bibliography{reference}{}
\bibliographystyle{aasjournal}


\end{CJK}
\end{document}